\documentclass[a4paper,11pt]{article}
\usepackage{pos}

\title{Superhumps and their Relation to the Disk Instability Model}
\ShortTitle{Superhumps}

\author*[a]{Daisaku Nogami}

\affiliation[a]{Department of Astronomy, Graduate Schoold of Science, Kyoto University,\\
  Kitashirakawa Oiwake-cho, Sakyo-ku, Kyoto 606-8502, Japan}

\emailAdd{nogami@kusastro.kyoto-u.ac.jp}

\abstract{Since the discovery of superhumps in 1974, these photometric
modulations have provided a crucial observational window into disk
instabilities in cataclysmic variable stars, particularly the tidal
instability associated with the 3:1 resonance. Over the past few decades,
extensive time-resolved photometry has revealed a rich diversity of
superhump-related phenomena, including delayed superhump development,
early superhumps in WZ Sge-type dwarf novae, systematic stage A-B-C
evolution, negative superhumps, and superhumps observed in related systems
such as intermediate polars and AM CVn stars.

In this invited review, we summarize key observational advances since
the establishment of the thermal-tidal instability framework, discuss
their theoretical interpretations within the disk instability model,
and highlight remaining open problems. These developments have been
driven by coordinated networks of amateur observers, wide-field robotic
surveys, and continuous high-precision space-based photometry from
Kepler and TESS. Together, they demonstrate that superhumps remain
a powerful probe of disk dynamics, binary parameters, and the interplay
between thermal, tidal, and geometric effects in accretion disks.}

\FullConference{87th Fujihara Seminar: The 50th Anniversary Workshop of the Disk
Instability Model in Compact Binary Stars (DIM50TH2025)\\
22-26 September 2025\\
Tomakomai, Japan\\}

\begin{document}
\maketitle

\section{Introduction}

Cataclysmic variable stars (CVs) consist of a late-type main sequence star
(secondary star) and a white dwarf (primary star) surrounded by an accretion
disk formed by gas transferred from the secondary star (\cite{war95book} for
a review).  Dwarf novae (DNe) are a class of CVs that undergo outbursts caused
by the thermal instability \cite{osa74DNmodel}.

Superhumps were first discovered in a dwarf nova VW Hyi by Vogt
\cite{vog74vwhyi} in 1974 and have been studied extensively by many researchers
since then. Superhumps are observed only during superoutbursts and are
characterized by their period typically a few percent longer than the
orbital period and the small amplitude of typically 0.1--0.3 mag
\cite{war85suuma}.  They are visible even in pole-on systems, crucially
indicating that they are non-geometrical in origin.  The memorable book
by Warner \cite{war95book} describes almost all of the phenomena observed
in cataclysmic variable stars in the literature by 1995, including
superoutbursts and superhumps in SU UMa stars.

The theoretical explanation for superhumps lies in the beat phenomenon between
the orbital motion and the precession of an eccentric disk \cite{osa89suuma}.
This eccentricity is caused by the tidal instability related to the 3:1
resonance in the accretion disk. This mechanism is central to the definition
of SU UMa-type dwarf novae.

The review by Osaki \cite{osa96review} summarizes the thermal-tidal
instability (TTI) model for dwarf nova outbursts which had been established
by 1996 in a framework connecting normal outbursts due solely to the thermal
instability and superoutbursts due to a conbination of thermal and tidal
instabilities (see also \cite{osa26dim50}).
The most important figure in this review is shown in Fig. 1.  The behavior
of these systems is largely governed by the mass transfer rate
($\dot{M}_{\rm tr}$) and the orbital period ($P_{\rm orb}$):
\begin{itemize}
    \item \textbf{Nova-likes / Permanent superhumpers:} High $\dot{M}_{\rm tr}$
          leads to stable disks.  The latter is tidally unstable and always
          shows superhumps.
    \item \textbf{Z Cam / SS Cyg:} Intermediate/low $\dot{M}_{\rm tr}$ results in
          systems dominated by thermal instability.  Z Cam stars are around
          the border line of the thermal instabiliy and sometimes show a
          standstill
    \item \textbf{ER UMa / SU UMa / WZ Sge:} Short $P_{\rm orb}$ (low mass ratio
          $q$) allows the disk to reach the 3:1 resonance, triggering tidal
          instability.  Larger $\dot{M}_{\rm tr}$ systems, namely ER UMa $>$
          SU UMa > WZ Sge, have shorter superoutburst-recurrence cycle.
\end{itemize}

\begin{figure}[ht]
  \begin{center}
  \includegraphics[height=5cm]{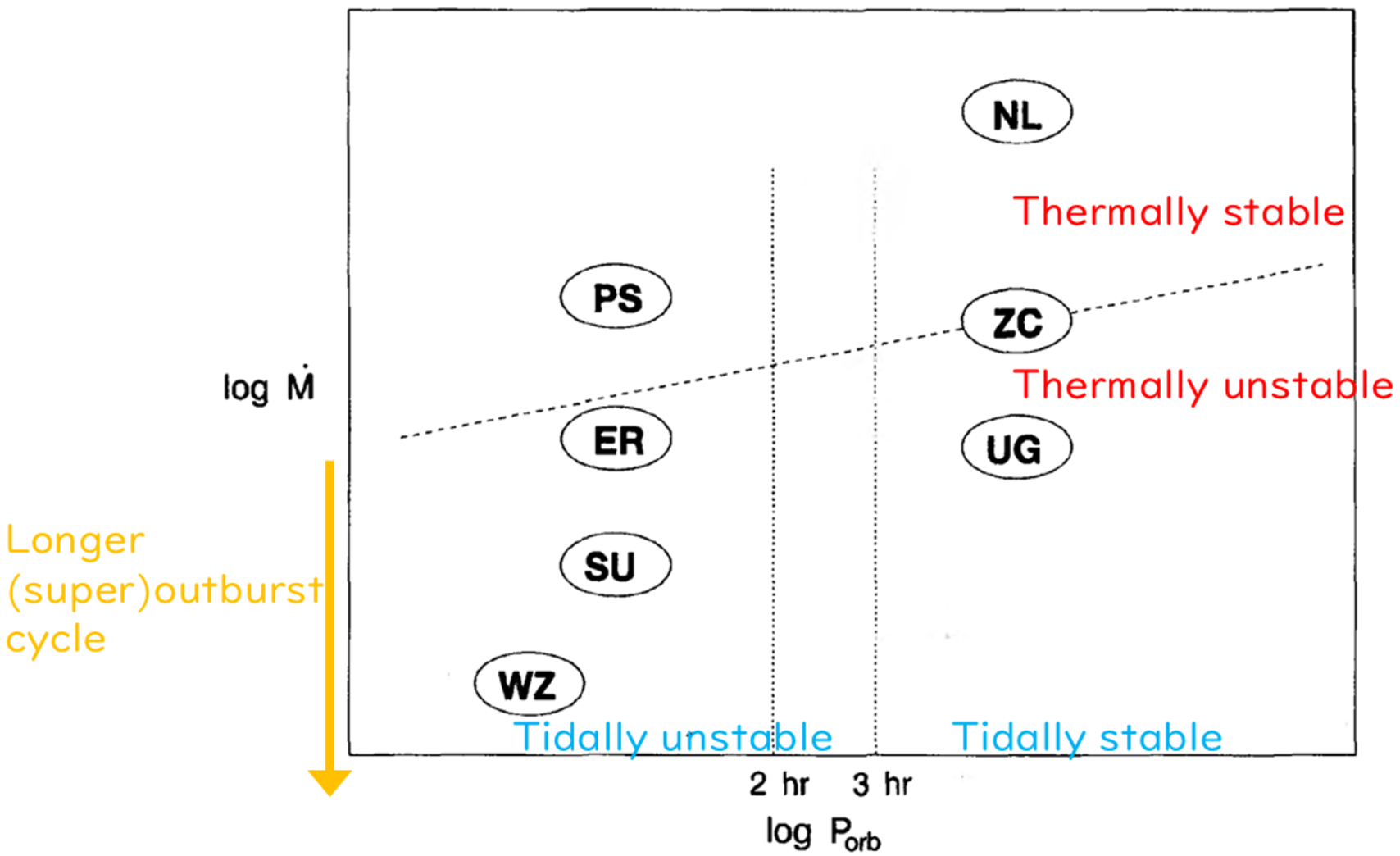}
  \end{center}
  \caption{$\dot{M}$-$P_{\rm orb}$ diagram.  The acronyms of NL, ZC,
    UG, PS, ER, SU and WZ present nova-likes, Z Cam-type stars,
    U Gem(SS Cyg)-type stars, permanent superhumpers, ER UMa-stars,
    SU UMa-stars and WZ Sge-type stars.  CVs in the upper and lower regions
    separated by the oblique line are thermally stable and unstable,
    respectively.  CVs in the right and left regions are tidally stable
    and unstable, respectively.  Systems with lower mass transfer rates have
    longer (super)outburst-recurrence cycles.  This figure was created based
    on Figure 3 in \cite{osa96review}.}
\end{figure}

In section 2, we summarize recent observational results of superhumps in
CVs, theoretical interpretation and remaining problems.  A brief summary is
put in section 3.

\section{Observations and Theoretical Explanations of Superhumps and related
phenomena}

This section examines several observational results concerning superhumps
in recent years that are difficult to explain by theories up to 1996, and
present their current theoretical interpretations and remaining problems.

\subsection{Superhump Evolution}

While earlier studies (e.g., \cite{pat93v603aql}) analyzed
the superhump evolution, long-term coverage was often insufficient to 
systematically discuss that evolution.  More recent systematic studies
(\cite{Pdot}, and its series of papers), however, have identified three
distinct stages of superhump evolution:
\begin{enumerate}
    \item \textbf{Stage A:} Characterized by a constant long period and
          growing amplitude.
    \item \textbf{Stage B:} The superhump period becomes shorter initially,
          which then lengthens while the amplitude decays in most cases.
          This represents the propagation of the eccentricity wave to the
          inner part of the disk.
    \item \textbf{Stage C:} Characterized by a constant short period.
          This stage typically appears around the end of the superoutburst
          and is accompanied by an decrease of the decline rate.
\end{enumerate}
An example is shown in figure 2.

\begin{figure}[ht]
  \begin{center}
  \includegraphics[height=9cm]{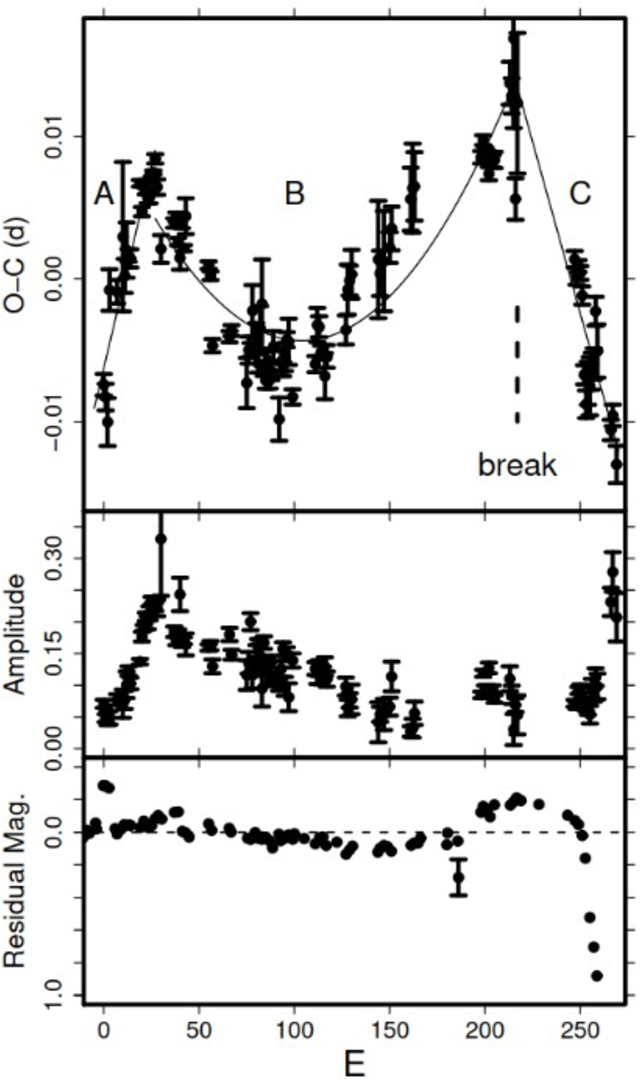}
  \end{center}
  \caption{(top) O-C diagram of the superhump maximum timing, (middle)
    superhump amplitude, and (bottom) superhump amplitude, (bottom) superoutburst
    light curve after subtraction of a linear decay trend.  The horizontal axis
    represents the superhump cycle ($E$) starting from the first observed
    superhump.  This figure is Figure 3 in \cite{Pdot}.}
\end{figure}

This course of superhump evolution is explained by the
DIM \cite{osa13v344lyrv1504cyg}.  The stage A is the growing period of
the superhumps as the eccentric deformation remains at the 3:1 resonance 
radius.  The period of the stage-A superhump thus represents the precession
rate at the 3:1 resonance radius.  During the stage B, the eccentricity wave
probagetes to the inner part of the disk.  The period of the stage-B
superhump represents the overall precession rate of the disk.  However, the
remaining problem is how the lengthening of the stage-B superhump period is
explained.  Note that some systems show constancy or even shortening of the
stage-B superhump period.  Furthermore, there are problems regarding the
stage C: why does the sudden transition from the stage B to C occur, and
why is the period of the stage-C superhump constant?

\subsection{A New Method for Mass Ratio Estimation}
Kato \& Osaki \cite{kat13qfromstageA} proposed that the mass ratio
($q \equiv M_{\rm 2}/M_{\rm 1}$) can be rather accurately estimated by the
excess of the stage-A superhump period and orbital period
($\epsilon^* \equiv (P_{\rm SH} - P_{\rm orb})/P_{\rm orb}$).
This idea is based on the formulation of the dynamical precession rate given by
Hirose \& Osaki \cite{hir90SHexcess} and the hypothesis that the stage-A
superhump period is determined by the precession rate at the 3:1 resonance
radius \cite{osa13v344lyrv1504cyg}.  Kato \& Osaki \cite{kat13qfromstageA}
gave an approximate formula of $q$ in the range of $0.025 \le q \le 0.394$:
\begin{equation}
q = -0.0016 + 2.60\epsilon^* + 3.33(\epsilon^*)^2 + 79.0(\epsilon^*)^3,
\end{equation}
where the maximum error in $q$ is estimated to be 0.0004.  Kato \&
Osaki\cite{kat13qfromstageA} checked the validity of this method by comparison
of the $q$ value by this method and that by the eclipse analysis (see also
\cite{kat22NewYear}).

While the mass ratio is a basic parameter of binaries, accurate estimation of
it has been difficult, since it requires observation of the eclipse profile
or radial velocity of the secondary star in CVs in quiescence.  In contrast,
we can estimate the mass ratio by observations of the early superhumps and
stage-A superhumps in dwarf novae in superoutburst, since the early superhump
period is almost equal to the orbital period \cite{ish02wzsgeletter}.  The new
and easy method has contributed to increases of the number of DNe with an
accurately measured mass ratio.

\subsection{Standstill and superoutburst in NY Ser}
NY Ser is a so-called ``in-the-gap" SU UMa star with a superhump period of
2.5 hr and a supercycle of about 100 days \cite{nog98nyser}.  Two standstills
were found in the 2018 light curve of this star \cite{kat19nyser}.

\begin{figure}[ht]
  \begin{center}
  \includegraphics[height=4cm]{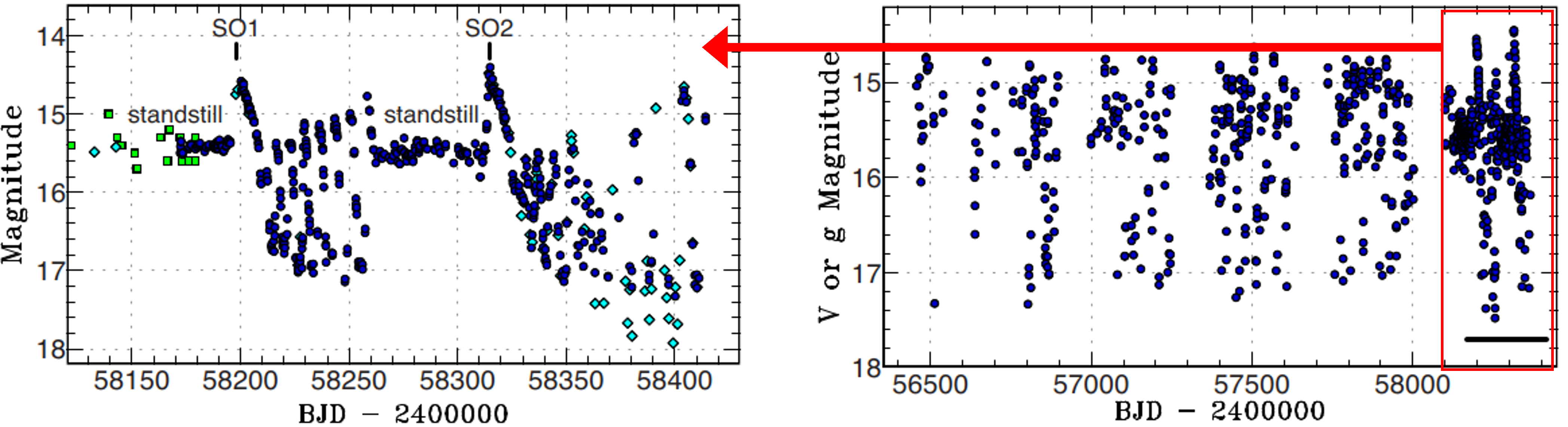}
  \end{center}
  \caption{(right) ASAS-SN light curve in $V$ and $g$ bands of NY Ser
    from 2015 to 2018. (left) Enlarged light curve in 2018.  Two superoutbursts
    occurred directly following standstills. This figure was created based on
    Figures 1 and 2 in \cite{kat19nyser}.}
\end{figure}

Both superoutbursts in 2018 rose directly from the immediately prior
standstills.  The second standstill clearly started during the decay from the
prior normal outburst, as often observed in Z Cam stars.  While no superhumps
were observed during the standstills, superhump evolution apparently began
at a very early phase of the second superoutburst rise from the standstill
(see E-figure 4 in \cite{kat19nyser}).

These observed features suggest that 1) the mass transfer rate in NY Ser is
close to the limit for the thermally stable disk\footnote{A problem remains
unsolved that why this star has such a high mass transfer rate even in the
period gap.}, 2) the 3:1 resonance was excited as the disk expanded to the
resonance radius during standstills, and 3) a superoutburst is triggered by
the tidal instability.

Related to NY Ser, we also note that RZ LMi, an ER UMa-type dwarf nova near
the period minimum was observed to behave like a permanent superhumper
intermittently (Kato et al.\cite{kat16rzlmi}).

\subsection{Long period SU UMa stars}
Long $P_{\rm orb}$ SU UMa stars have been discovered so far, e.g. TU Men
($P_{\rm orb}$ = 2.8 hr \cite{sto84tumen}).  Some of them have been proved
to have an evolved secondary star, e.g. CRTS J035905.9+175034
\cite{lit18crtsj0359}, and the relationship between orbital period and mass
ratio, assuming the secondary star is a main-sequence star, does not apply
to these stars.

However, Kato \& Vanmunster \cite{kat23sdssj0940} discovered a long
$P_{\rm orb}$ SU UMa star, SDSS J094002.56+274942.0 with a $P_{\rm SH}$
of 4.38 hr.  The superhumps had fully evolved by 6 days from the superoutburst
maximum.  Shallow eclipses were observed during the superoutburst with
the orbital period of 3.92 hr.  Using $P_{\rm orb}$, $P_{\rm SH}$ and
eclipses, a mass ratio of 0.39(3) and an inclination of 70.5(5)$\deg$ were
obtained.  The Gaia parallax and 2MASS observations support an idea of
a main-sequence secondary star in this system, and this idea can explain
ellipsoidal modulations observed during quiescence by ZTF and ATLAS (see
\cite{kra10sdssj0940}).

Numerical simulations have indicated the upper limit of the mass ratio for
development of the eccentricity by the 3:1 resonance is 0.25 \cite{whi88tidal}
or 0.33 \cite{mur00SHprecession}.  This discovery of such a high mass-ratio
SU UMa star, however, suggests that the eccentricity can grow under a weak
tidal effect and would request a reanalysis of the  basis of the 3:1
resonance.

\subsection{Early Superhumps in WZ Sge stars}
Early superhumps are observed exclusively in the very early phase of
WZ Sge-type superoutbursts and are regarded as one of the criteria for
classification as a WZ Sge-type dwarf nova \cite{kat15wzsge,tam26dim50}.
They exhibit double-peaked shapes and a period nearly equal to $P_{\rm orb}$.
The amplitude of the early superhumps depends on the inclination, suggesting
that early superhumps are geometrical phenomena (see e.g.
\cite{tam22pnvj0044}).

\begin{figure}[ht]
  \begin{center}
  \includegraphics[height=7cm]{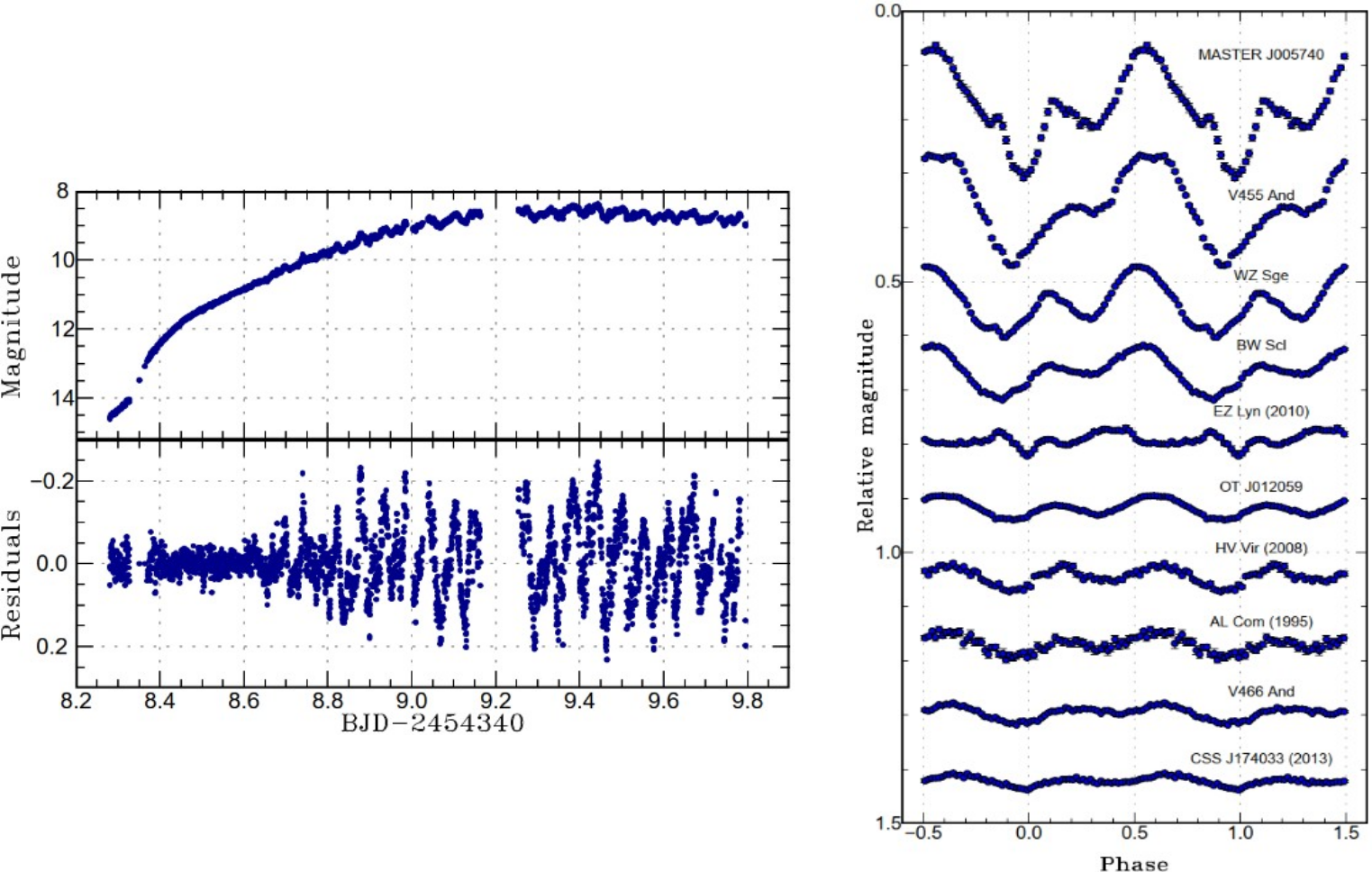}
  \end{center}
  \caption{(upperleft) Light curve of two days around the superoutburst
    maximum in a WZ Sge star V455 And in 2007.  (lowerleft) Its detrended one.
    The evolution of the early superhumps in a very short time ($<$ one day)
    was clearly observed. (right) Folded superhump profiles in different
    objects and superoutburst.  All have double-peaked shapes, though the
    details of these profiles are slightly different. This figure was
    created based on Figures 11 and 12 in \cite{kat15wzsge}.}
\end{figure}

In the TTI model, Osaki \& Meyer \cite{osa02wzsgehump} explained that the
early superhump appears due to the 2:1 resonance \cite{lin79lowqdisk,
lub91SHa}.  The 2:1 resonance radius is smaller than the tidal truncation
radius only in systems having a very small mass ratio ($q < 0.08$), and
the growth rate of the 2:1 resonance is much larger than the 3:1 resonance.
These features naturally explain that early superhumps are observed only
in WZ Sge stars before the (ordinary) superhumps appear.  Uemura et al.
\cite{uem12ESHrecon} developed a method to reconstruct the vertically
extended disk by modelling multi-color light curves of early superhumps
under an assumption of self-occultation (for an application of this
method, e.g. \cite{tam22pnvj0044}).

\subsection{Long period WZ Sge stars}
An outburst of ASASSN-16eg was first observed in 1996 \cite{wak17asassn16eg}.
The overall light curve shown in Fig. 5(a) seems truly a type-C superoutburst
(a superoutburst followed by only one short rebrightening) of WZ Sge-type DNe.
Both of early superhumps and ordinary superhumps were observed (Fig. 5), and
the periods of the early superhump and ordinary superhump were 109 and 115 min.
The mass ratio was estimated to be $q = 0.166(2)$ by using the period excess
method of the early superhump introduced above.  This mass ratio is normal
for its orbital period.  Wakamatsu et al.\cite{wak17asassn16eg} listed other
SU UMa-type dwarf novae which showed WZ Sge-like superoutbursts so far.

\begin{figure}[ht]
  \begin{center}
  \includegraphics[height=7cm]{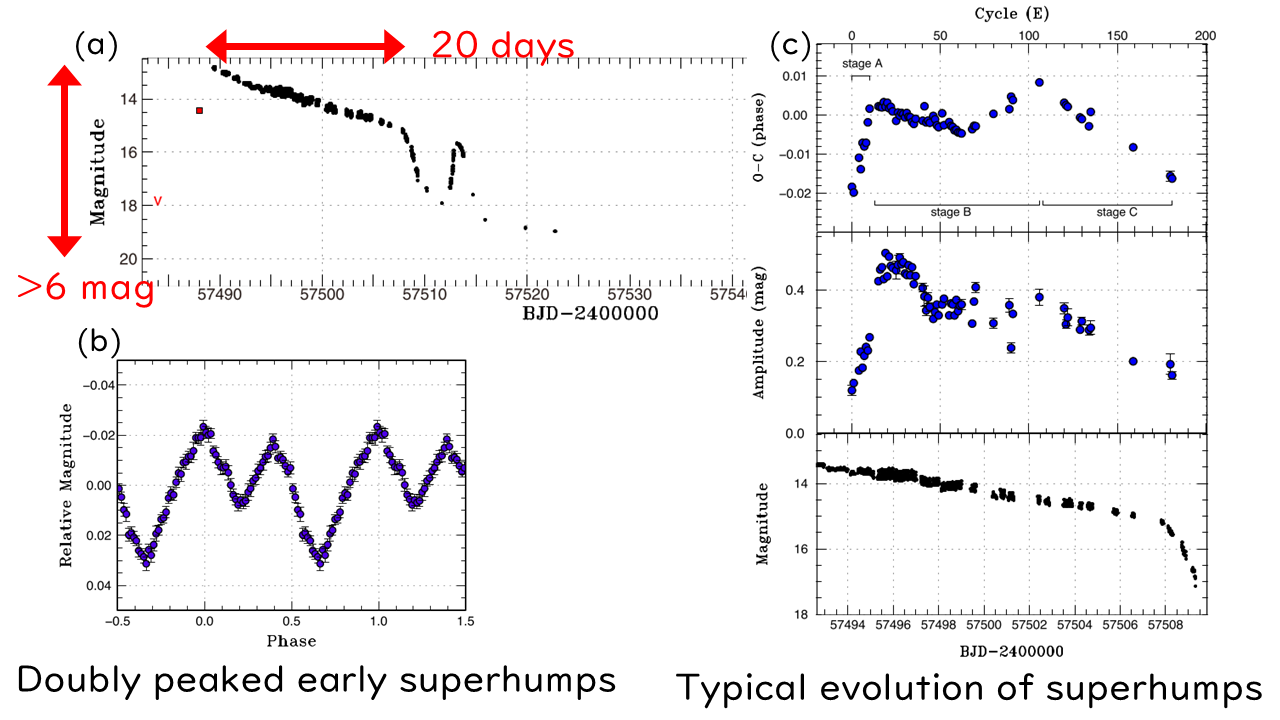}
  \end{center}
  \caption{(a) Long-term light curve of ASASSN-16eg during the
     superoutburst in 2016, which is a typical one of WZ Sge stars.
     (b) Folded light curve of early superhumps having a double-peaked
     shape.  (c) O-C diagram of the superhump maximum timing (upper),
     amplitude of the superhump, and (c) light curve of the superoutburst
     having a common horizontal axis.  This figure was created based on
     Figures 1, 2 and 3 in \cite{wak17asassn16eg}.}
\end{figure}

While the 2:1 resonance radius exceeds the tidal truncation radius,
can the TTI model explain WZ Sge-type superoutbursts in such "high-$q$"
systems?  This is still an open question.  However, such long period
"WZ Sge stars" commonly have long superoutburst cycle lengths over 2
years and large outburst amplitude over 6 mag.  This suggests that
these stars have very low mass transfer rate for their orbital periods.
In this case, the mass and angular momentum stored in the disk by the
onset of an outburst may be enough for the disk radius to exceed the
tidal truncation radius, to reach the 2:1 resonance radius around the
superoutburst maximum.  Confirmation by theoretical works and numerical
simulations are awaited.  This topic may be related to that cool gases
beyond the Roche-lobe of the white dwarf were detected during a
superoutburst in HT Cas \cite{neu20htcas}.

\subsection{Negative Superhumps}
Negative superhumps have a period slightly shorter than $P_{\rm orb}$
in contrast to (ordinary) superhumps (e.g. \cite{har95v503cyg}), and
they have been observed in various types of cataclysmic variables, such
as old novae (e.g. \cite{bru23TESSsuperhump}), nova-likes (e.g.
\cite{ver25dwcnc}), dwarf novae (e.g. \cite{jos25TESSsuperhump}),
AM CVn stars (e.g. \cite{sol21hplib}), and so on.  An example is
exhibited in figure 6.  Negative superhumps have the following features:
1) observed in CVs regardless of $P_{\rm orb}$ and $q$, 2) observed
regardless of the state, namely, quiescence, outburst, and standstill,
3) co-existence with (ordinary) superhumps, and 4) unpredictable timings
of appearance and disappearance.

\begin{figure}[ht]
  \begin{center}
  \includegraphics[height=6cm]{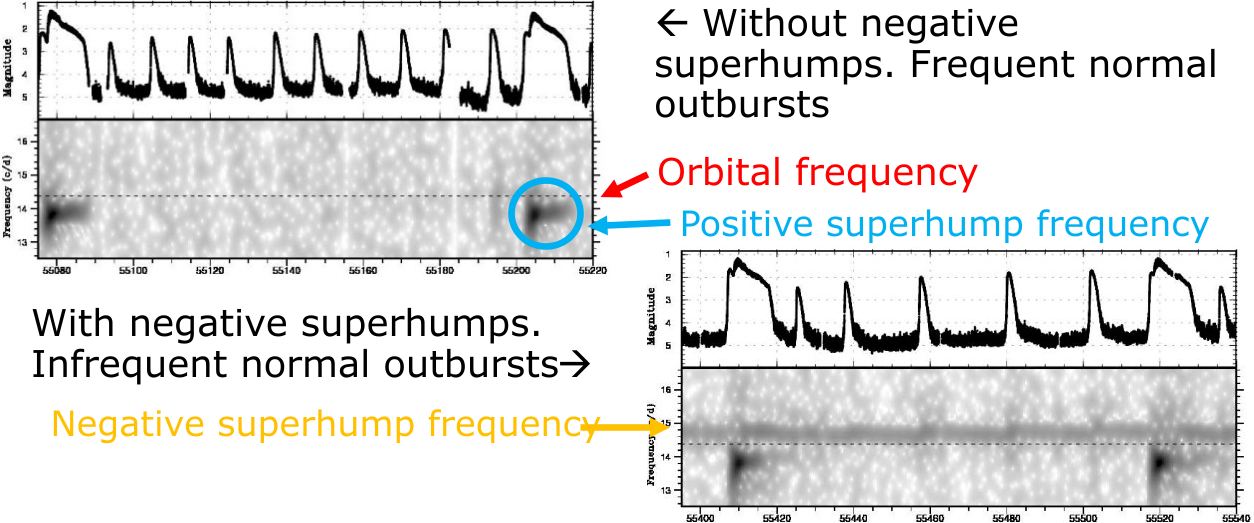}
  \end{center}
  \caption{(upperleft) Kepler light curve of an SU UMa star V1504 Cyg
          and its dynamic power spectra.  Only ordinary superhumps were observed
          during the superoutbursts.  (lowerright) The same, but in the the
          term when negative superhumps were observed as well as ordinary
          superhumps.  Normal outbursts occurred less frequently in this term.
          This figure was created based on Figure 3 in \cite{osa13v1504cygKepler}.}
\end{figure}

Theoretically, a tilted accretion disk is known to show retrograde precession
(e.g. \cite{pap83nonplanardisk}).  The period of the negative
superhump is explained by the synodic period between the orbital motion and
the retrograde precession of the disk (e.g. \cite{mon10disktilt}).
The difference of the normal outburst frequency is attributed to the difference
of the impact point of the mass stream on the disk \cite{osa13v1504cygKepler}.
Since disk expansion leads to an increase in the negative superhump frequency
\cite{lar98XBprecession}, variations of the negative superhump period provides
insight into disk expansion/shrinkage.  Recently, a new subgroup of dwarf novae
has been established, called "IW And"-type (\cite{sim11zcampaign,
szk13iwandv513cas, kat19iwand}), which are characterized by the light curves
of standstills with oscillations and outbursts directly rising from the standstill.
Numerical simulations trying to reproduce the IW And-type behavior by a combination
of the thermal instability and tilted disk has been performed \cite{kim20iwand,
kim26dim50}.

\section{Superhumps in related objects}

\subsection{Superhumps in an intermediate polar CC Scl}
Intermediate polars (IPs) have an accretion disk, the inner part of which is
truncated by strong magnetic fields on the white dwarf (\cite{pat94ipreview}
for a review of IPs,).  Some of DNe below the period gap have been proposed or
identified to be intermediate polars.  CC Scl is an example, whose $P_{\rm orb}$,
$P_{\rm SH}$, and spin period were estimated to be 1.4 hr \cite{che01ECCV,
tap04CTCV}, 1.44 hr \cite{wou12ccscl, kat15ccscl}, and 389.5 s \cite{pai24ccscl}.

\begin{figure}[ht]
  \begin{center}
  \includegraphics[height=7cm]{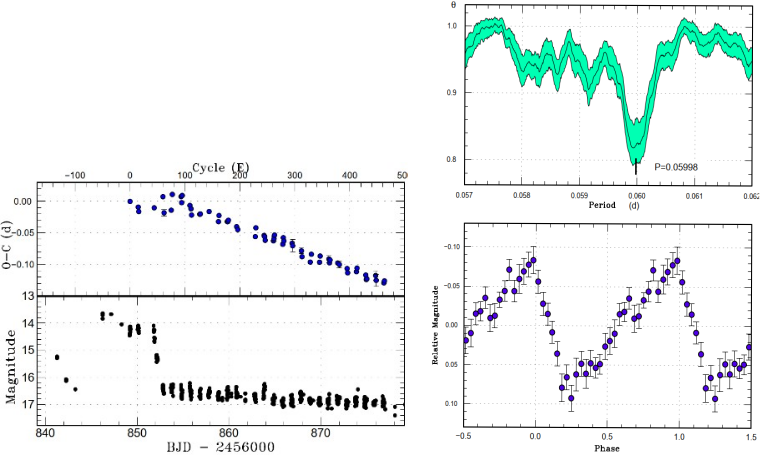}
  \end{center}
  \caption{(left) Light curve of the precursor and superoutburst
          in CC Scl in 2014 (lower) and $O-C$ diagram of the superhump
          maximum timing in the same horizontal axis (upper).
          (right) $\Theta$ diagram of a PDM analysis for superhumps
          (upper) and superhump light curve folded by the resultant
          period.
          This figure was created based on Figures 1 and 2 in
          \cite{kat15ccscl}.}
\end{figure}

The whole light curve of the superoutburst in CC Scl mimics that of
ordinary SU UMa stars (figure 7), which suggests that the TTI model
is applicable to IPs if the mass and angular momentum are stored,
enough for the disk radius to reach the 3:1 resonance radius, before
the onset of a precursor \cite{kat15ccscl}.  The duration of the
superoutburst is, however, 8 days at most, slightly shorter than
a typical value of SU UMa stars, which may be due to truncation
of the inner part of the accretion disk \cite{wou12ccscl}.  Although
the stage-A superhumps were missed, the stage-B and -C superhumps
were continuously observed.  The transition from the stage B to C
occurred about three days after the rapid decay from the superoutburst,
while this transition usually occurs before the rapid decay in ordinary
SU UMa stars.  This implies that the stage C superhumps originate
from the outermost part of the accretion disk \cite{kat15ccscl}.

\subsection{Superhumps in AM CVn Stars}
AM CVn stars are a kind of CVs but have a He star or a white dwarf as
the secondary (\cite{sol10amcvnreview, ram26dim50} for reviews of AM CVn
stars).  Superhumps and negative superhumps have been observed also in
AM CVn stars, and the behavior of them is basically the same as that of
H-rich CVs in a variety of types: nova-likes including permanent superhumpers,
Z Cam stars, ER UMa stars and SU UMa stars.  Transitions between the
states in some of AM CVn stars have been observed (e.g. \cite{pat00v803cen,
hon13crboo}).

\begin{figure}[ht]
  \begin{center}
  \includegraphics[height=4.5cm]{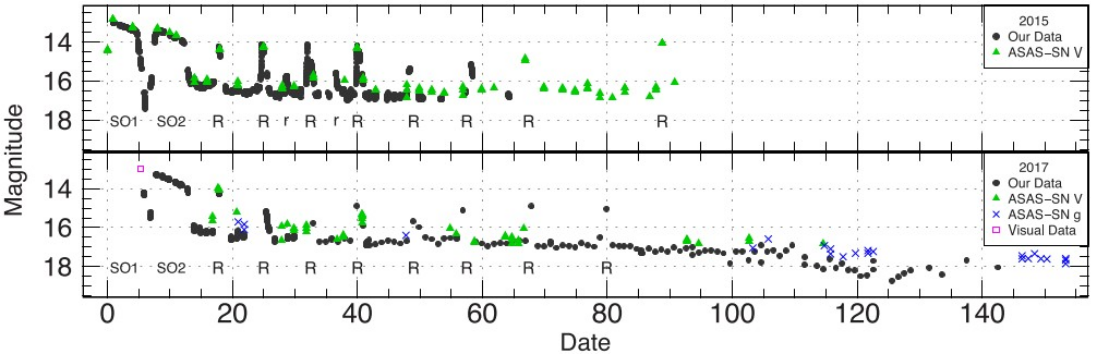}
  \end{center}
  \caption{Light curves of the 2015 and 2017 superoutbursts in NSV 1440
          in the upper and lower panels.  These mimic the type-E
          superoutburst (double superoutburst divided by a dip) and
          rebrightenings in WZ Sge stars.  This figure is Figures 1
          in \cite{iso19nsv1440}.}
\end{figure}

We here introduce an AM CVn star NSV1440.  Two superoutbursts were
observed in 2015 and 2017 \cite{iso19nsv1440}, and this star is the
first AM CVn star showing light curves analogous to those of type-E
superoutbursts and rebrightenings observed in H-rich WZ Sge stars
(see \cite{kat15wzsge}).  The early and stage-A superhump periods
in the 2015 superoutburst were 36.33(1) min and 36.98(3) min,
respectively.  The mass ratio calculated by using the period-excess
method was $q = 0.045(2)$.  The orbital period of this star was
photometrically measured to be $P_{\rm orb} = 36.56(3)$ min
\cite{aun25amcvns}.  By substituting this orbital period, the
period-excess method gives $q = 0.029(3)$.  In either case, NSV 1440
is certainly in the very late stage of the CV evolution, like WZ Sge
stars showing type-E superoutbursts.  This confirms that the DIM,
originally developed for H-rich disks, is also applicable to He-rich
disks.  Recently, analyses of such outbursts in six long-$P_{\rm orb}$
AM CVn stars clarified that all outbursts have very similar features,
especially in the long fading tail following the rebrightening phase
\cite{koj26longPorbAMCVns}.

\subsection{Superhumps in black hole X-ray binaries}
As the similarity between the outburst light curves of soft X-ray transients
and those of WZ Sge-type dwarf novae has been pointed out (e.g.
\cite{kuu00wzsgeSXT}), superhump-like optical variations have been
reported in several black hole X-ray binaries, e.g.
KV UMa \cite{wag01j1118,mcc01j1118spec,uem02j1118}. During the 2018
outburst of the black hole X-ray binary MAXI J1820+070, a clear
transition from stage A to stage B superhumps was observed for the
first time \cite{iij21asassn18ey}. The characteristics of the $O-C$
diagram of superhump maxima and the amplitude evolution closely resemble
those seen in SU UMa-type dwarf novae.

The mass ratio was estimated to be 0.066(1) using the period-excess method,
based on the stage-A superhump period and the orbital period
\cite{tor19maxij1820}. The black hole mass was estimated to be 7.3-10.8
$M_{\odot}$, consistent with the value of $8.48^{+0.79}_{-0.72} \ M_{\odot}$
derived from the rotational velocity. Furthermore, the superhump stage
transition occurred about 5 days earlier than the transition from the
hard to soft X-ray state. This behavior can be naturally interpreted
in terms of inward propagation of enhanced viscosity and the associated
viscous timescale \cite{iij21asassn18ey}.

These results indicate that the TTI model is applicable to outbursts
in black hole X-ray binaries, despite the presence of strong gravity
and intense X-ray irradiation. However, the large superhump amplitude
of approximately 0.8 mag and the absence of detectable color variation
during the superhump phase observed in this system remain unresolved
issues.

\section{Conclusion}

The disk instability model, supplemented by geometric effects, has evolved
into a remarkably successful framework for interpreting the diverse
phenomenology of superhumps in cataclysmic variables. Systematic observations
have established the stage A-B-C evolution of superhumps, enabled a new and
practical method for estimating binary mass ratios, and revealed unexpected
behaviors such as standstill-triggered superoutbursts, high-mass-ratio SU UMa
stars, long-period WZ Sge-like systems, and the widespread occurrence of
negative superhumps.

At the same time, these observations expose important theoretical challenges.
The origin of period changes during stage B, the abrupt transition to stage C,
the growth of eccentricity at unexpectedly high mass ratios, and
the excitation of the 2:1 resonance in long-period systems all require
further investigation through analytical work and numerical simulations.
The detection of superhumps in intermediate polars and AM CVn stars further
demonstrates that the essential physics of the disk instability model is
applicable across a wide range of disk compositions, magnetic environments,
and evolutionary stages.

The progress reviewed here underscores the unique role of superhumps as a
diagnostic of accretion-disk structure and dynamics. Insights gained from
cataclysmic variables provide a valuable foundation for understanding
accretion phenomena on longer timescales and larger spatial scales,
including those in X-ray binaries and active galactic nuclei.

\acknowledgments

This work is partly supported by JSPS KAKENHI Grant Number 24H00248.
We thank the amateur observers who have provided valuable data to VSNET,
AAVSO, and other organizations. Furthermore, the availability of
ground-based survey data and data from space telescopes Kepler and TESS
has greatly contributed as the foundation for the research introduced here.

\bibliographystyle{JHEP}
\bibliography{cvs}

\newcommand{\noop}[1]{}

\providecommand{\href}[2]{#2}\begingroup\raggedright\begin{thebibliography}{10}

\bibitem{war95book}
B.~Warner, \emph{Cataclysmic Variable Stars}, Cambridge University Press
  (1995).

\bibitem{osa74DNmodel}
Y.~{Osaki}, \emph{An accretion model for the outbursts of {U Geminorum} stars},
 PASJ, 26, 429 (1974).

\bibitem{vog74vwhyi}
N.~{Vogt}, \emph{Photometric study of the dwarf nova {VW Hydri}}, A\&A, 36, 369
(1974).

\bibitem{war85suuma}
B.~Warner, \emph{Stars that {GO} hump in the night -- the {SU UMa} stars},  in
  \emph{Interacting Binaries}, P.P.~Eggleton and J.E.~Pringle, eds., p.~367,
  1985.

\bibitem{osa89suuma}
Y.~{Osaki}, \emph{A model for the superoutburst phenomenon of {SU Ursae
  Majoris} stars}, PASJ, 41, 1005 (1989).

\bibitem{osa96review}
Y.~{Osaki}, \emph{Dwarf-nova outbursts}, PASP, 108, 39 (1996).

\bibitem{osa26dim50}
Y.~{Osaki}, \emph{A brief history of the disk instability model in cataclysmic
  variable stars}, in proceedings of "The 50th Anniversary Workshop of
  the Disk Instability Model in Compact Binary Stars" PoS(DIM50th2025)002
  (2026).

\bibitem{pat93v603aql}
J.~Patterson, G.~Thomas, D.R.~Skillman and M.~Diaz, \emph{The 1991 {V603
  Aquilae} campaign -- superhumps and $P$-dots}, ApJS, 86, 235 (1993).

\bibitem{Pdot}
T.~{Kato}, A.~{Imada}, M.~{Uemura}, D.~{Nogami}, H.~{Maehara}, R.~{Ishioka}
  et~al., \emph{Survey of period variations of superhumps in {SU UMa}-type
  dwarf novae}, PASJ, 61, S395 (2009).

\bibitem{osa13v344lyrv1504cyg}
Y.~{Osaki} and T.~{Kato}, \emph{Study of superoutbursts
  and superhumps in {SU UMa} stars by the {Kepler} light curves of {V344 Lyrae}
  and {V1504 Cygni}}, PASJ, 65, 95 (2013).

\bibitem{kat13qfromstageA}
T.~{Kato} and Y.~{Osaki}, \emph{New method to estimate binary mass ratios
  by using superhumps}, PASJ, 65, 115 (2013).

\bibitem{hir90SHexcess}
M.~{Hirose} and Y.~{Osaki}, \emph{Hydrodynamic simulations of accretion disks
  in cataclysmic variables -- superhump phenomenon in {SU UMa} stars},
  PASJ, 42, 135 (1990).

\bibitem{kat22NewYear}
T.~{Kato}, \emph{Evolution of short-period cataclysmic variables: implications
  from eclipse modeling and stage a superhump method (with new year\'s gift)},
  VSOLJ Variable Star Bulletin, 89 (2022).

\bibitem{ish02wzsgeletter}
R.~Ishioka, M.~Uemura, K.~Matsumoto, H.~Ohashi, T.~Kato, G.~Masi et~al.,
  \emph{First detection of the growing humps at the rapidly rising stage of
  dwarf novae {AL Com} and {WZ Sge}}, A\&A, 381, L41 (2002).

\bibitem{nog98nyser}
D.~Nogami, T.~Kato, H.~Baba and S.~Masuda, \emph{Discovery of the first
  in-the-gap {SU UMa}-type dwarf nova, {NY Serpentis} (={PG 1510$+$234})},
  PASJ, 50, L1 (1998).

\bibitem{kat19nyser}
T.~{Kato}, E.P.~{Pavlenko}, N.V.~{Pit}, K.A.~{Antonyuk}, J.V.~{Babina},
  A.V.~{Baklanov} et~al., \emph{Discovery of standstills in the SU UMa-type
  dwarf nova NY Serpentis}, PASJ, 71, L1 (2019).

\bibitem{kat16rzlmi}
T.~{Kato}, R.~{Ishioka}, K.~{Isogai}, M.~{Kimura}, A.~{Imada}, I.~{Miller}
  et~al., \emph{{RZ Leonis Minoris} bridging between {ER Ursae Majoris}-type
  dwarf nova and nova-like system}, PASJ, 68, 107 (2016).

\bibitem{sto84tumen}
B.~{Stolz} and R.~{Schoembs}, \emph{The {SU UMa} star {TU Mensae}}, A\&A,
  132, 187 (1984).

\bibitem{lit18crtsj0359}
C.~{Littlefield}, P.~{Gamavich}, M.~{Kennedy}, P.~{Szkody} and Z.~{Dai},
  \emph{A comprehensive K2 and ground-based study of CRTS J035905.9+175034, an
  eclipsing su uma system with a large mass ratio}, AJ, 155, 232 (2018).

\bibitem{kat23sdssj0940}
T.~{Kato} and T.~{Vanmunster}, \emph{SDSS J094002.56+274942.0: an SU UMa star
  with an orbital period of 3.92 hours and an apparently unevolved secondary},
  VSOLJ Variable Star Bulletin, 114 (2023).

\bibitem{kra10sdssj0940}
T.~{Kraijci} and P.~{Wils}, \emph{SDSS J094002.56+274942.0: an SU UMa star
  with an orbital period of 3.92 hours and an apparently unevolved secondary},
  Journal of AAVSO, 38, 33 (2010).

\bibitem{whi88tidal}
R.~Whitehurst, \emph{Numerical simulations of accretion disks. I - superhumps -
  a tidal phenomenon of accretion disks}, MNRAS, 232, 35 (1988).

\bibitem{mur00SHprecession}
J.R.~{Murray}, \emph{The precession of eccentric discs in close binaries},
  MNRAS, 314, L1 (2000).

\bibitem{kat15wzsge}
T.~{Kato}, \emph{{WZ Sge}-type dwarf novae}, PASJ, 67, 108 (2015).

\bibitem{tam26dim50}
Y.~{Tampo}, \emph{WZ Sge-type dwarf novae: an extreme laboratory in cataclysmic
  variables}, in proceedings of "The 50th Anniversary Workshop of
  the Disk Instability Model in Compact Binary Stars" PoS(DIM50th2025)014 (2026).

\bibitem{tam22pnvj0044}
Y.~{Tampo}, K.~{Isogai}, N.~{Kojiguchi}, M.~{Uemura}, T.~{Kato}, T.~{Tordai}
  et~al., \emph{PNV J00444033+4113068: Early superhumps with 0.7 mag amplitude
  and non-red color}, PASJ, 74, 1287 (2022).

\bibitem{osa02wzsgehump}
Y.~{Osaki} and F.~{Meyer}, \emph{Early humps in {WZ Sge} stars}, A\&A, 383, 574
  (2002).

\bibitem{lin79lowqdisk}
D.N.C.~Lin and J.~Papaloizou, \emph{Tidal torques on accretion discs in binary
  systems with extreme mass ratios}, MNRAS, 186, 799 (1979).

\bibitem{lub91SHa}
S.H.~{Lubow}, \emph{A model for tidally driven eccentric instabilities in
  fluid disks}, ApJ, 381, 259 (1991).

\bibitem{uem12ESHrecon}
M.~{Uemura}, T.~{Kato}, T.~{Ohshima} and H.~{Maehara}, \emph{Reconstruction of
  the structure of accretion disks in dwarf novae from the multi-band light
  curves of early superhumps}, PASJ, 64, 92 (2012).

\bibitem{wak17asassn16eg}
Y.~{Wakamatsu}, K.~{Isogai}, M.~{Kimura}, T.~{Kato}, T.~{Vanmunster},
  G.~{Stone} et~al., \emph{{ASASSN-16eg}: New candidate for a long-period {WZ
  Sge}-type dwarf nova}, PASJ, 69, 89 (2017).

\bibitem{neu20htcas}
V.V.~{Neustroef} and S.V.~{Zharikov}, \emph{Voracious vortices in cataclysmic
  variables. II. evidence for the expansion of accretion disc material beyond
  the roche lobe of the accretor in HT Cassiopeia during its 2017
  superoutburst}, A\&A, 642, A100 (2020).

\bibitem{har95v503cyg}
D.~Harvey, D.R.~Skillman, J.~Patterson and F.A.~Ringwald, \emph{Superhumps in
  cataclysmic binaries. {V}. {V503 Cygni}}, PASP, 107, 551 (1995).

\bibitem{bru23TESSsuperhump}
A.~{Bruch}, \emph{TESS light curves of cataclysmic variables -II- superhumps in
  old novae and novalike variables}, MNRAS, 519, 352 (2023).

\bibitem{ver25dwcnc}
M.~{Veresvarska}, S.~{Scaringi}, C.~{Littlefield}, D.~{de Martino},
  C.~{Knigge}, J.~{Paice} et~al., \emph{DW Cnc: a micronova with a negative
  superhump and a flickering spin}, MNRAS, 539, 2424 (2025).

\bibitem{jos25TESSsuperhump}
A.~{Joshi}, C.~{Tappert}, M.~{Catelan}, L.~{Schmidtobreick} and M.~{Singh},
  \emph{A tale of three cataclysmic variables with distinct superhumps},
  A\&A, 702, A70 (2025).

\bibitem{sol21hplib}
S.~{Solanki}, T.~{Kupfer}, O.~{Blaes}, E.~{Breedt} and S.~{Scarringi},
  \emph{Periodicities in the K2 lightcurve of HP Librae}, MNRAS, 500,
  1222 (2021).

\bibitem{osa13v1504cygKepler}
Y.~{Osaki} and T.~{Kato}, \emph{The cause of the superoutburst in
  {SU UMa} stars is finally revealed by {Kepler} light curve
  of {V1504 Cygni}}, PASJ, 65, 50 (2013).

\bibitem{pap83nonplanardisk}
J.C.B.~{Papaloizou} and J.E.~{Pringle}, \emph{The time-dependence of non-planar
  accretion discs}, MNRAS, 202, 1181 (1983).

\bibitem{mon10disktilt}
M.M.~{Montgomery} and E.L.~{Martin}, \emph{A common source of accretion disk
  tilt}, ApJ, 722, 989 (2010).

\bibitem{lar98XBprecession}
J.~{Larwood}, \emph{On the precession of accretion discs in {X-ray} binaries},
  MNRAS, 299, L32 (1998).

\bibitem{sim11zcampaign}
M.~{Simonsen}, \emph{The Z CamPaign: Year 1}, Journal of AAVSO, 39,
  66 (2011).

\bibitem{szk13iwandv513cas}
P.~{Szkody}, M.~{Albright}, A.P.~{Linnell}, M.E.~{Everett}, R.~{McMillan},
  G.~{Saurage} et~al., \emph{A study of the unusual {Z Cam} systems {IW
  Andromedae} and {V513 Cassiopeia}}, PASP, 125, 1421 (2013).

\bibitem{kat19iwand}
T.~{Kato}, \emph{Three Z Camelopardalis-type dwarf novae exhibiting iw
  andromedae-type phenomenon}, PASJ, 71, 20 (2019).

\bibitem{kim20iwand}
M.~{Kimura}, Y.~{Osaki}, T.~{Kato} and S.~{Mineshige}, \emph{Thermal-viscous
  instability in tilted accretion disks: A possible application to IW
  Andromeda-type dwarf novae}, PASJ, 72, 22 (2020).

\bibitem{kim26dim50}
M.~{Kimura}, \emph{Could the thermal instability in the tilted disk be
  responsible for iw and phenomenon?}, in proceedings of "The 50th
  Anniversary Workshop of the Disk Instability Model in Compact Binary
  Stars" PoS(DIM50th2025)012 (2026).

\bibitem{pat94ipreview}
J.~Patterson, \emph{The {DQ Herculis} stars}, PASP, 106, 209 (1994).

\bibitem{che01ECCV}
A.~Chen, D.~O'Donoghue, R.S.~Stobie, D.~Kilkenny and B.~Warner,
  \emph{Cataclysmic variables in the {Edinburgh-Cape Blue Object Survey}},
  MNRAS, 325, 89 (2001).

\bibitem{tap04CTCV}
C.~{Tappert}, T.~{Augusteijn} and J.~{Maza}, \emph{{Cataclysmic variables from
  the Cal{\'a}n-Tololo Survey - I. Photometric periods}}, MNRAS, 354, 321
  (2004).

\bibitem{wou12ccscl}
P.A.~{Woudt}, B.~{Warner}, A.~{Gulbis}, R.~{Coppejans}, F.-J.~{Hambsch},
  A.P.~{Beardmore} et~al., \emph{{CC Sculptoris}: A superhumping intermediate
  polar}, MNRAS, 427, 1004 (2012).

\bibitem{kat15ccscl}
T.~{Kato}, F.-J.~{Hambsch}, A.~{Oksanen}, P.~{Starr} and A.~{Henden}, \emph{{CC
  Sculptoris}: Eclipsing {SU UMa}-type intermediate polar}, PASJ, 67, 3 (2015).

\bibitem{pai24ccscl}
J.A.~{Paice}, S.~{Scaringi}, N.~{Castro Segura}, A.~{Sahu}, K.~{I\l kiewicz},
  D.L.~{Coppejans} et~al., \emph{Evolution of spin in the intermediate polar CC
  Sculptoris}, MNRAS, 531, L82 (2024).

\bibitem{sol10amcvnreview}
J.~{Solheim}, \emph{{AM CVn} stars: Status and challenges}, PASP, 122, 1133
  (2010).

\bibitem{ram26dim50}
G.~{Ramsay}, \emph{Outbursts in ultra-compact AM CVn binaries}, in proceedings
  of "The 50th Anniversary Workshop of the Disk Instability Model in Compact
  Binary Stars" PoS(DIM50th2025)017 (2026).

\bibitem{pat00v803cen}
J.~Patterson, S.~Walker, J.~Kemp, D.~O'Donoghue, M.~Bos and R.~Stubbings,
  \emph{{V803 Centauri}: A helium-rich dwarf nova}, PASP, 112, 625 (2000).

\bibitem{hon13crboo}
R.K.~{Honeycutt}, B.R.~{Adams}, G.W.~{Turner}, J.W.~{Robertson}, E.M.~{Ost} and
  J.E.~{Maxwell}, \emph{Light curve of {CR Bootis} 1990--2012 from the
  {Indiana} long-term monitoring program}, PASP, 125, 126 (2013).

\bibitem{iso19nsv1440}
K.~{Isogai}, T.~{Kato}, B.~{Monard}, K.~{Kasai}, F.-J.~{Hambsch}, G.~{Myers}
  et~al., \emph{NSV 1440: first WZ Sge-type object in AM CVn stars and
  candidates}, PASJ 71, 48 (2019).

\bibitem{aun25amcvns}
A.~{Aungwerojwit}, B.T.~{G\"ansicke}, E.~{Breet}, S.~{Arjyotha}, J.J.~{Hermes},
  F.J.~{Hambsch} et~al., \emph{Follow-up on three poorly studied AM CVn stars},
  MNRAS, 537, 3078 (2025).

\bibitem{koj26longPorbAMCVns}
N.~{Kojiguchi}, K.~{Isogai}, Y.~{Tampo}, T.~{Kato}, H.~{Itoh}, B.~{Manard}
  et~al., \emph{Observational study of double superoutbursts in
  long-orbital-period AM CVn stars}, PASJ, 78, in press (2026).

\bibitem{kuu00wzsgeSXT}
E.~Kuulkers, \emph{{WZ Sge} stars/{TOADs} and (soft) {X}-ray transients: close
  encounters of the same kind}, New Astrnomy, 44, 27 (2000).

\bibitem{wag01j1118}
R.M.~Wagner, C.B.~Foltz, T.~Shahbaz, J.~Casares, P.A.~Charles, S.G.~Starrfield
  et~al., \emph{The halo black hole x-ray transient XTE J1118+480}, ApJ,
  556, 42 (2001).

\bibitem{mcc01j1118spec}
J.E.~McClintock, C.A.~Haswell, M.R.~Garcia, J.J.~Drake, R.I.~Hynes,
  H.L.~Marshall et~al., \emph{Complete and simultaneous spectral observations
  of the black hole x-ray nova XTE J1118+480}, ApJ, 555, 477 (2001).

\bibitem{uem02j1118}
M.~Uemura, T.~Kato, K.~Matsumoto, H.~Iwamatsu, R.~Ishioka, L.M.~Cook et~al.,
  \emph{Optical observations of {XTE J1118$+$480} during the 2000 outburst},
  PASJ, 54, 285 (2002).

\bibitem{iij21asassn18ey}
K.~{Niijima}, M.~{Kimura}, T.~{Wakamatsu}, Y.~{Kato}, D.~{Nogami}, K.~{Isogai},
  N.~{Kojiguchi} et~al., \emph{Optical variability correlated with X-ray
  spectral transition in the black-hole transient ASASSN-18ey = MAXI
  J1820+070}, VSOLJ Variable Star Bulletin, 74 (2021).

\bibitem{tor19maxij1820}
M.A.P.~{Torres}, J.~{Vasares}, F.~{Jim\'enez-Ibarra}, T.~{Mu\"noz-Darias},
  M.~{Armas Padilla}, P.G.~{Jonker} et~al., \emph{Dynamical confirmation of a
  black hole in MAXI J1820+070}, ApJ, 882, 21 (2019).

\end{thebibliography}\endgroup



\end{document}